\documentstyle[epsfig]{mn2e}

\begin{document}
\title [Suppression of instabilities by halo] 
{Suppression of gravitational instabilities by dominant dark matter 
halo in low surface brightness galaxies}
\author[S. Ghosh  and C.J. Jog ]
       {Soumavo Ghosh$^{1}$\thanks{E-mail : soumavo@physics.iisc.ernet.in}, and
        Chanda J. Jog$^{1}$\thanks{E-mail : cjjog@physics.iisc.ernet.in}\\
$^1$   Department of Physics,
Indian Institute of Science, Bangalore 560012, India \\
}

\maketitle

\begin{abstract} 
The low surface brightness galaxies are gas-rich and yet have a low star formation rate, this is a well-known puzzle.
The spiral features in these galaxies are weak and difficult to trace, although this aspect has not been studied much.
These galaxies are known to be dominated by the dark matter halo from the innermost regions.
 Here we do a stability analysis for
the galactic disc of UGC 7321, a  low surface brightness, superthin galaxy, for
which the various observational input parameters are available. We show that the disc is stable against local, linear axisymmetric
and non-axisymmetric perturbations. The Toomre $Q$ parameter values are found to be large ($>> 1$) mainly due to the low disc surface density and
the high rotation velocity resulting due to the dominant dark matter halo, which could explain the observed low star formation rate.
For the stars-alone case, the disc shows finite swing amplification but the addition of dark matter halo suppresses that amplification almost completely. Even the inclusion of the low-dispersion gas which constitutes a high disc mass fraction 
does not help in causing swing amplification.
This can explain why these galaxies  do not show strong spiral features. 
Thus the dynamical effect of a halo that is dominant from inner regions can naturally explain why star formation and spiral features are largely suppressed
in low surface brightness galaxies, making these different from the high surface brightness galaxies.
\end{abstract}


\section{Introduction} 

It is well-known that the class of galaxies known as the Low Surface Brightness (LSB) galaxies are characterized by a low disc surface density (de Blok \& McGaugh 1996, de Blok, McGaugh \& Rubin 2001) and a low star formation rate (Impey \& Bothun 1997). 
Despite having a high 
mass fraction of the disc in gas, these do not appear to have had much star formation  and are thus
unevolved (Bothun, Impey \& McGaugh 1997).  This has been a long-standing basic puzzle about these galaxies. 
Many LSB galaxies do show spiral structure but it is fragmentary or incipient, and is extremely faint and difficult to trace
(Schombert et al. 1990,
McGaugh, Schombert \& Bothun 1995, also see Schombert, Maciel \& McGaugh 2011, Figs. 5 and 6).

The lack of star formation in the LSBs is not well-understood. One common reason suggested for this is the low surface density that lies below the onset for threshold (van der Hulst et al. 1993), or the lack of molecular gas which normally forms the site of star formation (O'Neil, Schinnerer \& Hofner 2003, Jog 2012). Further, these galaxies are seen in isolated regions (Mo, McGaugh \& Bothun 1994) or at the edges of voids (Rosenbaum et al. 2009).  Thus there is a lack of galaxy interaction that could have triggered
star formation (e.g., Mihos, McGaugh \& de Blok 1997) or spiral features (e.g., Noguchi 1987).

It is well-established in the literature that the LSB galaxies are dark matter dominated down to their central regions (Bothun et al. 1997,  de Blok \& McGaugh 1997, de Blok et al. 2001, Combes 2002, Banerjee, Matthews \& Jog 2010). We note that this is different from the dominance of dark matter at large radii as deduced from the flat rotation curves that is seen in all galaxies. In the LSBs, within the optical disc, the dark matter constitutes about 90\% of the total mass and the rest (10\%) is in baryons, while within the same radius, the two are comparable in the "normal" or high surface density (HSB) galaxies (e.g., de Blok et al. 2001, Jog 2012). It has sometimes been
claimed that the dark matter halo prevents star formation   though the details  are not well-understood. 

Early work using numerical simulations has indicated  (Mihos et al. 1997) that the LSBs are stable against the growth of global non-axisymmetric modes such as bars due to low surface density and large dark matter content. However, observations show that bars are common in LSBs and can be off-centred as in irregular galaxies (Matthews \& Gallagher 1997).  This needs to be followed up in future work. 
 The discrepancy could be partly due to different definitions of bars used in these papers.

Some of the dynamical properties of LSB galaxies such as stability against a bar mode 
(Mihos et al. 1997), and the superthin nature of a particular subset of LSB galaxies as arising due to a dense and compact halo (Banerjee \& Jog 2013), have been studied so far. However, the absence of strong spiral features in LSB galaxies has not drawn the much-deserved attention so far.

In this paper, we present a dynamical study of local axisymmetric and non-axisymmetric perturbations in a  LSB, superthin galaxy, UGC 7321,
for which the observational parameters are known. A previous study of this galaxy has revealed it is dark matter dominated at all radii starting from the innermost regions (Banerjee et al. 2010). Also the de-projected radial $HI$ surface density lies well below the critical $HI$ surface density for star formation for all radii (Uson \& Matthews 2003).
 Thus UGC 7321 is a suitable candidate to probe the effect of dark matter halo on both axisymmetric and non-axisymmetric local perturbations. We find that this galaxy is highly stable against both these mainly due to the low disc surface density and the high fraction of dark matter from the innermost regions which dominates its rotation curve.

We obtain the values for the $Q$ parameter for local stability first for gas-alone (Toomre 1964) and then for gravitationally coupled stars and gas as in Jog (1996).
The value of $Q < 1$ is taken to indicate onset of star formation (Kennicutt 1989). We also study the local non-axisymmetric linear perturbations using the idea of swing amplification as a mechanism for generating local spiral features (Goldreich \& Lynden-Bell 1965 - GLB hereafter, Toomre 1981). This amplification is temporary and arises as a mode swings past the radial direction, and occurs due a kinematic resonance between the epicyclic motion and the trailing sense of differential rotation in a disc, when supported by the self-gravity in the disc.
While using swing amplification as a tool, people often use some educative guess values for the  $Q$ parameter, but in this paper we calculate the actual values of $Q$ parameters obtained from the observed quantities and use these as input to study the local axisymmetric stability and also the swing amplification. 
 We show that in a galaxy like UGC 7321  the dark matter halo dominates in mass from the innermost regions, which increases the $Q$ values and hence suppresses star formation and swing amplification nearly completely.

Section 2  presents the details of the formulation of the problem, section 3 presents the details of input parameters, and the numerical solution and the results. Section 4 contains the
conclusions.

\section{Formulation of the Problem}
\subsection{Axisymmetric case}
A disc supported by rotation and random motion is stable to local, axisymmetric perturbations if the following $Q$ criterion is satisfied (Toomre 1964)
\begin{equation}
Q \: = \: \kappa c / \pi G \mu > 1 
\end{equation}
\noindent where $\kappa$ is local epicyclic frequency, $c$ is the one-dimensional random velocity dispersion and $\mu$ is the surface density of the disc. 

This was extended for a two-component disc consisting of gravitationally coupled stars and gas by Jog (1996), 
where the local stability parameter  $ Q_{s-g}$ for axisymmetric  two-fluid case is defined as 
\begin{equation}
[\frac{2{\pi}G k{\mu}_s}{{\kappa}^2+k^2c_s^2}+\frac{2{\pi}Gk{\mu}_g}{{\kappa}^2+k^2c_g^2}]_{ at \: k_{min}} \equiv \frac{2}{1+(Q_{s-g})^2}
\end{equation}
\noindent where  $k$ is the wavenumber of the perturbation and $k_{min}$ is the wavenumber at which it is hardest to stabilize the two-fluid system, and $c_s$ and $c_g$ denote the random velocity dispersion in stars and gas respectively. 
While this is difficult to solve analytically, a semi-analytical approach to solve this was given by Jog (1996) 
 which is summarised next.

Three dimensionless parameters are used, $Q_s$ and $Q_g$ the standard $Q$ parameters for local stability of  stars-alone and gas-alone cases respectively, and $\epsilon = \mu_g / (\mu_{g} + \mu_{s})$ the gas mass fraction in the disc.
In terms of the above three dimensionless parameters, the above condition reduces to 
\begin{eqnarray}
\frac{(1-\epsilon)}{l_{s-g}\{1+[Q_s^2(1-\epsilon)^2]/(4l_{s-g}^2)\}}+\frac{\epsilon}{l_{s-g}[1+Q_g^2{\epsilon}^2/(4l_{s-g}^2)]}\nonumber\\
\equiv \frac{2}{1+(Q_{s-g})^2}
\end{eqnarray}
\noindent where 
$l_{s-g}= [{\kappa}^2/{2{\pi}Gk_{min}({\mu}_s+{\mu}_g)}]$ is the dimensionless wavelength at which it is hardest to stabilize the two-fluid system. In analogy with the one-component case, the
 disc is shown to be stable, marginally stable or unstable  against axisymmetric perturbations depending on whether the $Q_{s-g}$ value is $> 1$,  = 1, or $ < 1$ respectively (Jog 1996).

A normal-mode linear perturbation analysis 
of the two-fluid (stars coupled with gas) system supported by random motion and rotation (see Jog \& Solomon 1984) showed that a radial mode $(k,{\omega})$ obeys the dispersion relation 
\begin{equation}
{\omega}^2(k)=\frac{1}{2}\{({\alpha}_s+{\alpha}_g)-[({\alpha}_s+{\alpha}_g)^2-4({\alpha}_s{\alpha}_g-{\beta}_s{\beta}_g)]^{1/2}\}, 
\end{equation}
\noindent where \\
${\alpha}_s = {\kappa}^2+k^2{c}_s^2-2{\pi}G k{\mu}_s, $            $ {\alpha}_g = {\kappa}^2+k^2{c}_g^2-2{\pi} G k{\mu}_g$ ,
\begin{equation}
{\beta}_s=2{\pi} G k{\mu}_s,    {\beta}_g=2{\pi} G k{\mu}_g
\end{equation}
\noindent  where $k$ is the wavenumber (=$ 2 \pi / \lambda$) where $\lambda$ is the wavelength
 and $\omega$ is the angular frequency of the perturbation.

For a given set of three dimensionless parameters introduced above
we numerically find the minimum of the dimensionless dispersion relation ${\omega}^2 (k) /\kappa^2$ function, 
and thus obtain the corresponding $k_{min}$.
To do this we vary the function between the wavelengths for the minima for the gas-alone and
stars-alone cases ($l_g = Q_g^2\epsilon^2/2$ and $l_s = Q_s^2(1-\epsilon)^2/2$) since the minimum of the dispersion relation will occur at a wavelength that lies between these two values (Jog \& Solomon 1984).  Then substituting this value of $k_{min}$ in equation (3) yields the value of $Q_{s-g}$ for a particular choice of values for the set $Q_s$, $Q_g$ and $\epsilon$.

\subsection{Non-axisymmetric case}
We consider a local, non-axisymmetric linear perturbation analysis of the galactic disc, where the perturbation is taken to be planar. The treatment follows as in GLB for a one-component disc, and for a two-fluid case in Jog (1992).
 The disc is taken to be thin for the sake of simplification of calculation as in Jog (1992).
We next introduce the sheared coordinates $(x',y',z',t')$ defined by
\begin{eqnarray}
x'=x,\: y'=y-2Axt,\: z'=z, \: t'=t
\end{eqnarray}
\noindent These were introduced by GLB to study the dynamics of a local patch in a differentially rotating disc. A trial solution of the form $exp[i(k_xx'+k_yy')]$ for the perturbed quantities including the density
$\delta \mu$ is introduced.
Next, we define ${\tau}$ (for a wavenumber $k_y\ne 0$):
\begin{equation}
\tau \equiv 2At'-k_x/k_y
\end{equation}
\noindent where $x$ is along the initial radial direction. In the sheared frame, $\tau$ is a measure of time and it has different zeroes for
each mode characterized by a different $k_x/k_y$ (Jog 1992). In each case, $\tau = 0$ when the 
wave-vector is along the radial direction. 
Further, $\delta \mu$ represents the density variation or amplification with time in the sheared frame
while in the non-sheared frame centred on the galactic centre, it gives the density for a mode of wavenumber $[k_y (1 + \tau^2)^{1/2}]$
that is sheared by an angle $\gamma = tan^{-1} \tau$ with respect to the radial direction, and $\mu$ is the unperturbed disc surface density.

Define ${\theta}_i$, the dimensionless perturbation surface density to be 
\begin{equation}
\theta_i={\delta \mu_i}/{\mu_{i}}
\end{equation}
Using the above trial solution and making use of the definition of ${\tau}$,  the linearized perturbation equations for the Euler equation, continuity equation and the joint Poisson equation combine to give the following two coupled differential equations which describe the evolution with $\tau$ in $\theta_i$ 
\begin{eqnarray}
~~~~~~~~~\left(
\frac{d^2\theta_i}{d \tau^2}
\right)-\left(
\frac{d\theta_i}{d\tau}
\right)\left(
\frac{2\tau}{1+\tau^2}
\right)\nonumber\\
~~~~~~~~~~~~~+\theta_i\left[
\frac{\kappa^2}{4A^2}+\frac{2B/A}{1+\tau^2}+\frac{k_y^2}{4A^2}(1+\tau^2){c_i^2}
\right]\nonumber\\
\!\!\!\!\!\!\!=(\mu_{s}\theta_s+\mu_{g}\theta_g)
\left(
\frac{\pi G k_y}{2A^2}\right)\!\!(1+\tau^2)^{1/2},
\end{eqnarray}
 where $i = s, g$ correspond to stars and gas respectively.
 The three terms
in the parentheses on the l.h.s describe the effect of the epicyclic motion, the sheared motion and the pressure term.

\noindent The analogous equation for the one-fluid (say stellar case) case is obtained by setting $\mu_g=0$ in equation (9):
\begin{eqnarray}
~~~~~~~~~~\left(
\frac{d^2\theta_s}{d \tau^2}
\right)\!\!-\!\!\left(
\frac{d\theta_s}{d\tau}
\right)\!\!\!\left(
\frac{2\tau}{1+\tau^2}
\right)\nonumber\\
~~~~~~~~~~~+\theta_s\!\!\left[
\frac{\kappa^2}{4A^2}+\frac{2B/A}{1+\tau^2}+\frac{k_y^2}{4A^2}(1+\tau^2){c_s^2}\right]
\nonumber\\
\!\!\!\!\!\!\!\!\!\!-\mu_{s}\left(
\frac{\pi G k_y}{2A^2}
\right)(1+\tau^2)^{1/2}
=0,
\end{eqnarray}
Following the two-fluid approach by Jog (1992), we introduce the dimensionless parameters as: $Q_s, Q_g$ the $Q$ factors for stars-alone and gas-alone respectively, $\epsilon$, the gas mass fraction in the disc, $\eta={2A}/{\Omega}$
a measure of differential rotation in the disc, and $X = \lambda_y / \lambda_{crit} $ where $\lambda_{crit} = 4 \pi^2 G (\mu_{s} + \mu_{g})/ \kappa^2$.

Using these dimensionless parameters the above coupled equations reduce to
\begin{eqnarray}
\left(
\frac{d^2\theta_s}{d \tau^2}
\right)-
\left(
\frac{d\theta_s}{d\tau}
\right)
\left(
\frac{2\tau}{1+\tau^2}
\right)~~~~~~~~~~~~\nonumber\\
~~+{\theta_s}\left[
{\xi^2}+\frac{2(\eta-2)}{\eta(1+\tau^2)}+\frac{(1+\tau^2)Q_s^2(1-\epsilon)^2\xi^2}{4X^2}
\right]\nonumber\\
=\frac{\xi^2}{X}(1+\tau^2)^{1/2}[\theta_s(1-\epsilon)+\theta_g\epsilon]\\
\left(
\frac{d^2\theta_g}{d \tau^2}
\right)-
\left(
\frac{d\theta_g}{d\tau}
\right)
\left(
\frac{2\tau}{1+\tau^2}
\right)~~~~~~~~~\nonumber\\
~~+{\theta_s}\left[
{\xi^2}+\frac{2(\eta-2)}{\eta(1+\tau^2)}+\frac{(1+\tau^2)Q_g^2\epsilon^2\xi^2}{4X^2}
\right]\nonumber\\
=\frac{\xi^2}{X}(1+\tau^2)^{1/2}[\theta_s(1-\epsilon)+\theta_g\epsilon]
\end{eqnarray}
\noindent where ${\xi^2}={\kappa^2/4A^2}= 2(2-\eta)/\eta^2$\\
\noindent Similarly, the one-fluid analog for the stars-alone case is given by 
\begin{eqnarray}
\left(
\frac{d^2\theta_s}{d\tau^2}
\right)-
\left(
\frac{d\theta_s}{d\tau}
\right)
\left(
\frac{2\tau}{1+\tau^2}
\right)~~~~~~~~~~\nonumber\\
+{\theta_s}\left[
{\xi^2}+\frac{2(\eta-2)}{\eta(1+\tau^2)}+\frac{(1+\tau^2)Q_s^2\xi^2}{4X^2}-\frac{\xi^2}{X}(1+\tau^2)^{1/2}
\right]=0
\end{eqnarray}

To bring out the sole effect of non-axisymmetric perturbations, we solve the above equations in the special case when the system is stable against the axisymmetric perturbation.
The necessary condition for axisymmetric stability is (see Jog \& Solomon 1984)
\begin{equation}
\frac{(1-\epsilon)}{X'\{1+[Q_s^2(1-\epsilon)^2/4X'^2]\}}+\frac{\epsilon}{X'[1+(Q_g^2\epsilon^2/4X'^2)]} < 1
\end{equation}
\noindent $X'=\lambda_a/\lambda_{crit}$ and $\lambda_a$ is the wavelength of the axisymmetric perturbation.
Thus for each set of parameters, equations $(11)$ and $(12)$ are to be solved while ensuring that the inequality given by equation (14) is satisfied for all wavelengths ranging from $Q_g^2\epsilon^2/2$ to $Q_s^2(1-\epsilon)^2/2$.

\section{Numerical Solution }
\subsection{Input parameters}
For the stellar disc and halo parameters, we assume the values obtained observationally and by modelling respectively (Banerjee et al. 2010).
The central surface density of the stellar disc is obtained as  $50.2 M_{\odot} pc^{-2}$ and the value of the exponential disc scale length 
$R_d$ is  $2.1$ kpc. This gives a radial variation of the stellar disc surface density required to calculate the $Q_s$ parameters for the stellar disc.  Note that this is very low, nearly a factor of 12 smaller than the central value for the Galaxy (e.g., Narayan \& Jog 2002).
 For the dark matter halo, a pseudo-isothermal density profile is obtained (see eq. [19]), where the core radius, $R_c$, is obtained to be $2.5$ kpc and the core density $\rho_o$ is  $0.057 M_{\odot} pc^{-3}$.
The gas surface density as a function of radius is taken from Uson \& Matthews (2003, see Fig. 13 in that paper).

For the stars, the one dimensional velocity dispersion in the z-direction is taken to vary with radius as $c_s= (c_s)_0 \: exp(-R/2R_d)$ (Banerjee et al. 2010) where$(c_s)_0$ is  equal to $14.3 $ $km s^{-1}$ . For the solar neighbourhood, it is observationally found that, the ratio of $(c_s)_z$ to the radial dispersion is $\sim 0.5$ (e.g., Binney \& Tremaine 1987). Here we assume the same conversion factor for all radii in this galaxy.

Banerjee  et al. (2010) found that to get the best fit to the HI scale height data, a higher value of gas velocity dispersion is needed in the inner parts whereas a slightly lower value is required for the outer regions of UGC 7321. Here we assume the same values they used, thus imposing a small gradient in the gas velocity dispersion by letting it vary linearly between $9.5$ $kms^{-1}$ at $R=7$ kpc and $8$ $kms^{-1}$ at $R=12.6$ kpc.
\subsection{Solution of Equations and Results}
\subsubsection{Axisymmetric case} Based on the above input parameters,   we next calculate the  values of $Q_s$ and $Q_g$ at different radii for  UGC 7321. The value of $\kappa$ is obtained from observed rotation curve which already includes the effect of the dark matter halo and the gas.
For a given set of values of input parameters $Q_s$,$Q_g$ and $\epsilon$, we obtain the stability parameter $Q_{s-g}$ for a two-component disc (see the procedure outlined in Sec. 2.1).
The corresponding values are given in Table $1$.

It is clear from Table 1 that at all radii, 
the joint $Q_{s-g}$ values are well above $1$, implying that the galaxy is highly stable against the two-fluid axisymmetric perturbations.
For gas-alone the $Q$ parameter is $>>1$ due to the low gas density, and a high $\kappa$ which reflects the effect of the dominant halo
in setting the undisturbed rotational field. While the star-gas gravitational interaction does lower the two-component values, these are still $> 3$ at all radii.
 Thus as a whole the galaxy is stable all the way starting from the inner part up to the very outer region. This is opposite to the picture for a high surface brightness galaxy such as the Milky way. The joint $Q_{s-g}$ values in the inner part are close to $1$ in that case, indicating that the Galactic disc is close to onset of
 two-fluid axisymmetric instability (see Table 1 of Jog 1996).
Thus, it is the dark matter halo that is dominant from the inner regions in the LSB galaxy UGC 7321 studied here that helps to stabilize the disc against one-fluid as well as two-fluid local, axisymmetric perturbations.

Note that $Q < 1$ is routinely used as a criterion for the onset of star formation (Kennicutt 1989). Hence we can argue that the
high values of the $Q$ parameter even for a two-component case, as obtained from observational input parameters for UGC 7321 
indicates  
why the LSB galaxies have a low star formation rate.
 Thus, our result confirms the previous result on the lack of star formation  by
van der Hulst et al. (1993), and Uson \& Matthews (2003), and puts it on a more firm quantitative ground.
\nopagebreak
\begin{table}
\centering
  \begin{minipage}{140mm}
   \caption{$Q$ values at different radii of UGC 7321}
\begin{tabular}{lllll}
$R/R_d$ & $\epsilon$, gas fraction & $Q_s$ & $Q_g$ & $Q_{s-g}$ \\
\\
1.0 & 0.33 & 4.6 & 9.7 & 3.8 \\
1.5 & 0.39 & 4.4 & 6.4 & 3.4 \\
2.0 & 0.42 & 4.0 & 5.3 & 3.0\\
2.5 & 0.50 & 4.3& 5.3 & 3.0 \\
3.0 & 0.54 & 5.2 & 6.5 & 3.5 \\
3.5 & 0.50 & 6.1 & 11.6 & 4.4 \\
4.0 & 0.50 & 7.1 & 15.4 & 5.2 \\
\end{tabular}
\end{minipage}
\end{table}

\subsubsection{Non-axisymmetric case}

We first consider a stars-alone galactic  disc for which the epicyclic frequency $\kappa$ is obtained theoretically. The definition of $\kappa$
at the mid-plane for a general potential $\Phi$ is defined by (Binney \& Tremaine 1987):
\begin{equation}
{\kappa}^2=
\left[
\frac {{\partial}^2 \Phi}{\partial R^2}+\frac{3}{R}\frac{{\partial}\Phi}{\partial R}
\right]_{z=0}
\end{equation}
The potential $\Phi(R,0)$ for an exponential disc in the equatorial plane is given by (Binney \& Tremaine 1987, eq. 2.168)
\begin{equation}
{\Phi}(R,0)=-{\pi}G{\Sigma}_0R[I_0(y)K_1(y)-I_1(y)K_0(y)]
\end{equation}
\noindent where ${\Sigma}_0$ is the disc central surface density, y is the dimensionless quantity given by $y=R/2R_d$, $R$ being galactocentric radius and $R_d$ being disc scale-length and $I_n$ and $K_n$$(n=0,1)$ are the modified Bessel function of the first and second kind respectively.
For this choice of  potential, equation (15) gives
\begin{eqnarray}
{\kappa}^2_{disc}=\frac{{\pi}G{\Sigma}_0}{R_d}[4I_0K_0-2I_1K_1+2y(I_1K_0-I_0K_1)]
\end{eqnarray}
\noindent Using this expression, we calculate the $Q_s$ values for the disc treated as a stars-alone case.
At $R=3R_D$ which is used for illustrative purposes here, the resulting value of $Q_s$ = 1.2.

In presence of a dark matter halo, the unperturbed rotational velocity and hence $\kappa$ and hence $Q$ are higher. This is particularly 
expected to be true for LSB galaxies like UGC 7321  when the dark matter dominates from the inner regions.
This fact is normally not realized since $\kappa$ is obtained using the observed rotation curve which already includes the effect of the halo.

To illustrate the effect of considering a disc in halo, we theoretically calculate the net epicyclic frequency.
In this case, the net ${\kappa}^2$ is given by adding the values for the disc alone and the halo in quadrature
\begin{eqnarray}
{\kappa}^2_{net} = {\kappa}^2_{disc} + {\kappa}^2_{halo}
\end{eqnarray}
A spherical, pseudo-isothermal halo is found to give the best fit while modelling the halo parameters of UGC 7321 using the $HI$ scale height data and observed rotation curve as simultaneous constraints (Banerjee et al. 2010). To keep parity with that work here also we take an pseudo-isothermal halo as characterized by
\begin{eqnarray}
{\rho}(r)=\frac{{\rho}_0}{(1+r^2/R_c^2)}
\end{eqnarray}
\noindent\ where ${\rho}_0$ 
 is the core density and $R_c$ is the core radius of the dark matter halo.
Using this profile we solve the Poisson equation  to obtain the potential and then converting it into standard galactic cylindrical co-ordinates $(R,\phi,z)$, the final expression for potential due to the dark matter halo is
\begin{eqnarray}
{\phi}_{halo}=(4{\pi}{\rho}_0 R_c^2) 
\bigg[\frac{1}{2}\log(R_c^2+R^2+z^2)
+(\frac{R_c}{(R^2+z^2)^{1/2}})\nonumber\\
\tan^{-1}(\frac{(R^2+z^2)^{1/2}}{R_c})-1\bigg]~~~~~~~~~~~~~~~~~~~~
\end{eqnarray}
\noindent  Using this in equation (15) yields
\begin{eqnarray}
~~~~~~~~~~~~{\kappa_{halo}^2}= 4{\pi}G {\rho}_0 R_c^2
\bigg[
\frac{2}{R_c^2+R^2}+\nonumber\\
\;\;\;\;\;\;\;\;\;\;\left(
\frac{R_c^2}{R^2}
\right)
\frac{1}{R_c^2+R^2}-\frac{R_c}{R^3}{\tan}^{-1}(\frac{R}{R_c})
\bigg],
\end{eqnarray}
The value of $\kappa_{net}$ is obtained by combining the disc and the halo contributions (equations 17, 21) as in equation (18).
At $R=3R_D$ which is used for illustrative purposes here, using this value of $\kappa_{net}$ the  resulting value of $Q_s$ = 5.0. Note that this is much higher than for
the stars-alone disc ($Q_s = 1.2$) and this leads to suppression of swing amplification in presence of a halo as shown in Fig. 1.

Though the rotation curve for UGC 7321 is not perfectly flat, instead slightly rising right after $2R_d$, but for sake of simplicity of calculation we assume a flat rotation curve for from $2R_d$ for this galaxy. Hence we set $\eta=1$ and $\xi^2=2$ in equations (11-13) for the cases discussed here.

Here we consider the analysis  at $R=3R_d$.
We first consider the stars-alone case
and then solve by adding the effect of dark matter halo as described above. This scheme is used to check how strong is the effect of dark matter halo on the swing amplification (see Fig. 1).

For a given set of  parameters, we have to solve the second-order linear differential equation $(13)$ for the one-fluid case. We treat it as a set of two coupled first order linear differential equation in $\theta_s$ and ${d\theta_s}/{d\tau}$ and then solve them numerically in an iterative fashion using Fourth order Runge-Kutta with the initial values at $\tau_{ini}$, the initial value of $\tau$. Following GLB, the possible set of initial values for these two variables at 
$\tau_{ini}$ are (1,0) and (0,1).
As the values of $(\theta_s)_{max}$ and MAF$(=(\theta_s)_{max}/(\theta_s)_{ini})$ depend crucially on the choice of $\tau_{ini}$ (Jog 1992), we varied both the $\tau_{ini}$ and initial values of the variables simultaneously to get the maximum swing amplification possible for that given set of parameters.

A typical solution starting at large negative $\tau$ values shows an oscillatory behaviour due to the dominance of the pressure term.
As the mode goes past the radial direction the shearing term and the epicyclic term are in a temporary kinematical resonance, the duration of which is enhanced due to the self-gravity of the perturbation, this gives rise to the swing amplification as the mode swings past the radial direction (or $\tau =0$ in the sheared frame). At large $\tau$ values the pressure term dominates again. This explanation (see GLB, Toomre 1981) describes the schematic behaviour of the solution seen in Figure 1 (top panel). 
\begin{figure}
\centering
\includegraphics[height=2.4in,width=3.2in]{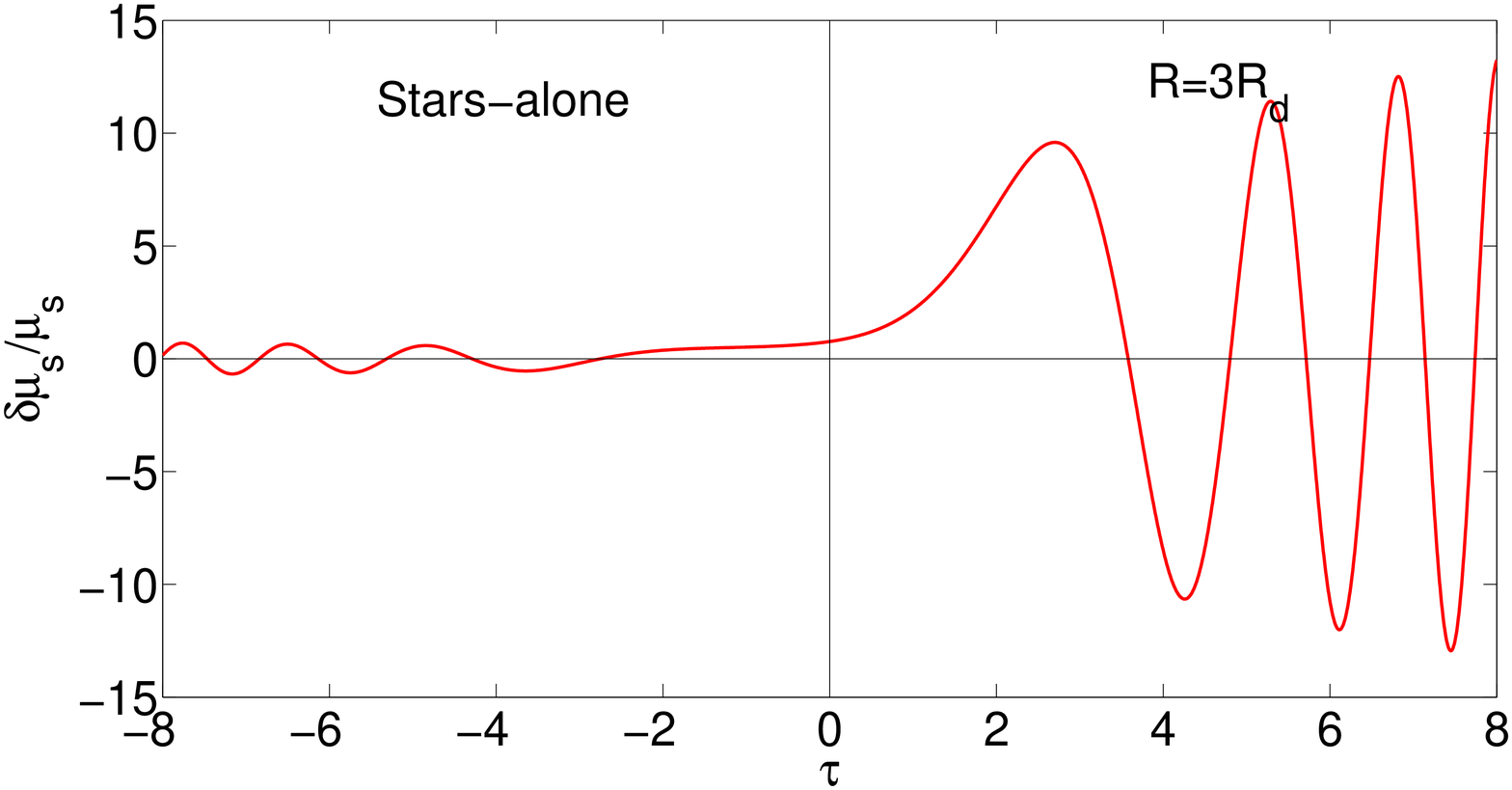}
\medskip
\includegraphics[height=2.4in,width=3.2in]{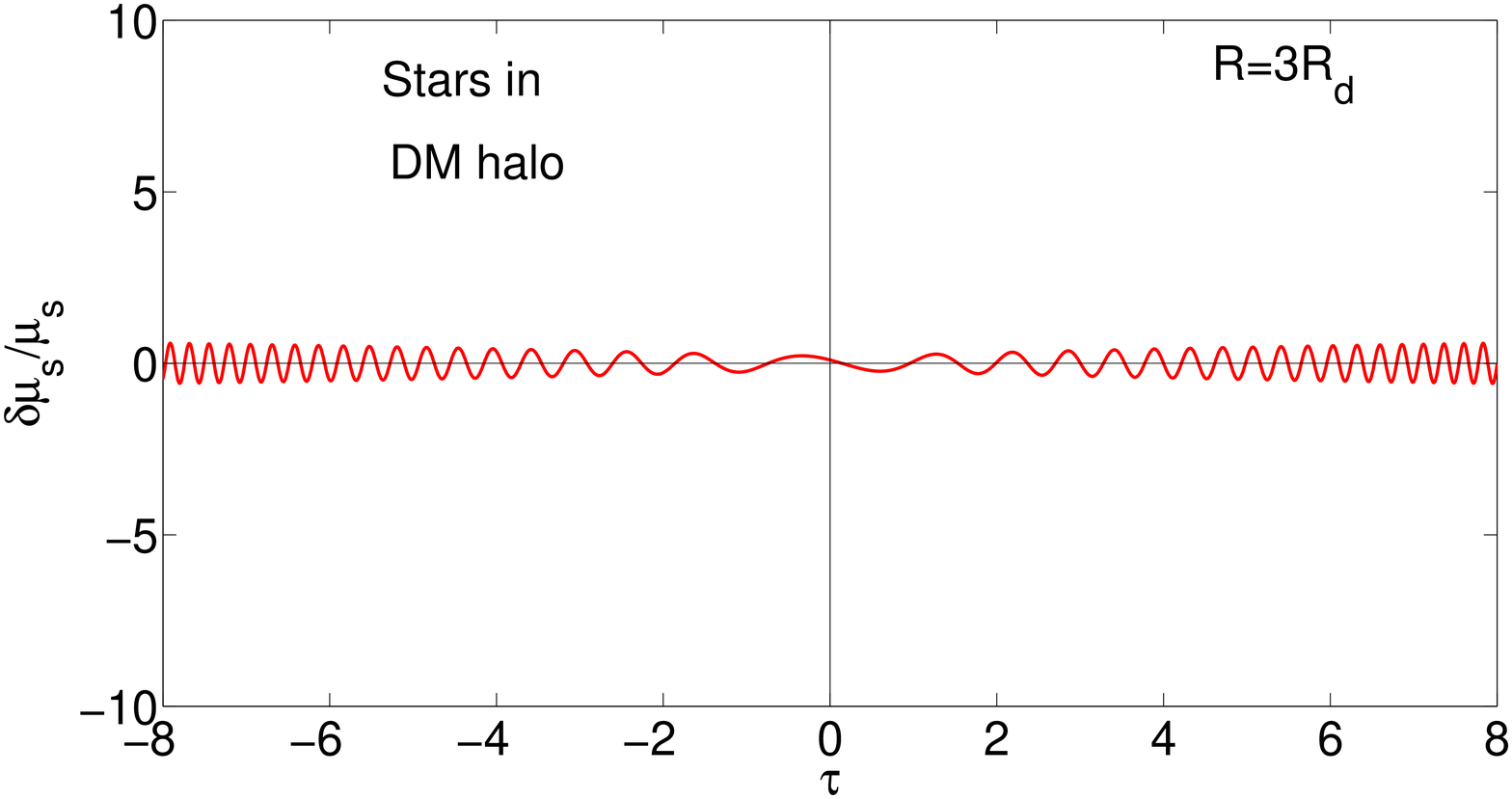}
\caption{Variation in $\theta_s=\delta\mu_s/\mu_s$, the ratio of the perturbation surface density to the unperturbed surface density,with $\tau$, dimensionless time in the sheared frame. The top panel shows 
the stars-alone case ($Q_s=1.2$, $\eta=1$, and $X=1$) at the radius $R= 3R_d$ and the lower panel shows
the case of stars in the dark matter halo  ($Q_s=5.0$, $\eta=1$, and $X=1$) at the same radius. The stars-alone disc shows a finite amplification, but when the stars are taken in the field of the dark matter halo, the amplification is almost completely damped. This is due to the higher rotational velocity and $Q$ values in presence of the dominant dark matter halo.}
\end{figure}
\noindent When the stars-alone disc is considered then the disc shows finite swing amplification, but the addition of the dark matter halo suppresses the swing amplification completely. Thus this galaxy will not exhibit strong spiral features.

For the sake of completeness, we next also
 include the gas and treat the galactic disc as a gravitationally coupled two-fluid system to make our approach more realistic.
Here while calculating the ${\kappa}$ we use the best fit of observed rotation curve of that galaxy obtained by Banerjee et al. (2010).
This includes the effect of dark matter halo as well as stars and gas in determining the undisturbed rotational velocity in the galactic disc. At $R=3R_D$, this gives $Q_s = 5.2$ and $ Q_g = 6.5$ (see Table 1). Note that this value of $Q_s$ obtained from the observed rotation curve is close to that obtained by treating a disc in a dark matter halo potential ($Q$=5.0), this shows the small change due to the inclusion of gas in the real case.

Given a set of parameter values, we are to solve  equations $(11)$ and $(12)$ simultaneously for the two-fluid case. These are treated as 
a set of four first order linear coupled differential equations in $\theta_s$ ,${d\theta_s}/{d\tau}$,$\theta_g$ and ${d\theta_g}/{d\tau}$ and then solve them numerically in a  way similar to what was done in Jog (1992).

\begin{figure}
\centering
\includegraphics[height=2.4in,width=3.2in]{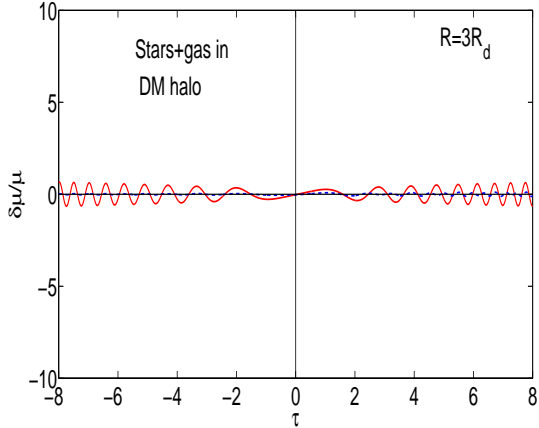}
\caption{Variation in $\theta=\delta\mu_i/\mu_i$, the ratio of the perturbation surface density to the unperturbed surface density,with $\tau$, dimensionless time in the sheared frame treating galactic disc as a two-fluid gravitationally coupled system consisting of stars and gas in the field of the dark matter halo at $R= 3 R_d$. $Q_s=5.2$, $Q_g=6.5$, $X=1$, $\eta=1$ and the gas fraction $\epsilon=0.54$.
Despite the high gas fraction, the addition of gas does not make any drastic change, hardly any swing amplification is seen for both stars and gas.}
\end{figure}

 From Fig. 2  it is clear that the inclusion of gas makes hardly any impression  as there is no sign of swing amplification in both stars and gas. This is in contrast to the strong effect of gas shown in a typical HSB galaxy (Jog 1992). This is because of the low disc surface density and the large dark matter content in UGC 7321, that prevents the supply of necessary disc self-gravity to amplify the non-axisymmetric dynamical seeds. This supports a similar conclusion by Mihos et al. (1997) for the bar instabilities.

As the rotation curve for UGC 7321 is not exactly flat, instead a slightly rising beyond $2R_d$ (see Fig.1 of Banerjee et al. 2010) 
then the value of $\eta$ will be different from $1$. We 
calculated the value of $\eta$ from observed rotation curve and found this to give $\eta =0.36$
 and re-did the above calculation for the two-fluid case. Even the consideration of actual $\eta$ value does not make any drastic change, hardly any swing amplification is seen for both stars and gas.
 
\section {Discussion}
 \subsection {Weak spiral structure in LSBs and its origin}
Some LSB galaxies do show spiral structure although in general it is  
 weak and difficult to trace (Section 1). However, in view 
of the analysis in the present paper which shows a suppression of 
instabilities, the surprise is that the LSB discs do show some 
structure at all albeit feeble. In this section, we try to address 
this interesting question in terms of the amplitude of spiral features, 
and 
the mechanisms that could give rise to the weak features.

A quantitative measurement by a modal analysis on lines of 
that done routinely for the HSB galaxies for $m=2$ (spiral arms or bars) 
or $m=1$ (lopsidedness) (Rix \& Zaritsky 1995, 
Bournaud et al. 2005) has not been done for the LSB galaxies. But a visual
inspection shows the spiral features in LSBs to be much fainter and 
fragmentary than in the HSB galaxies (McGaugh et al. 1995, Schombert 
et al. 2011). A modal analysis for the LSBs needs to be carried out to 
measure the Fourier amplitudes of the various modes quantitatively.

In this paper we have focused on swing amplification as a mechanism for 
generation of local spiral features. It has been shown (Toomre 1981, 
Sellwood \& Carlberg 1984) that this mechanism stops being effective 
for $Q > 2-2.5$ or $X > 3$ (where $X$ is the dimensionless wavelength (see Section 2.2). 
Beyond this range, strong swing amplification 
does not occur (Binney \& Tremaine 1987). However, this does not rule out
the possibility of spiral features occurring outside this range of 
parameters. The disc could still support occasional, weak spiral features.
This is because a differentially rotating disc is always susceptible to the growth of 
non-axisymmetric features due to the kinematic resonance between the 
epicyclic motion and the shear both of which have the same sense of 
motion. Even if the disc gravity is feeble, this resonance could last 
for about dynamical time-scale ($\sim 10^ 8$ yr), and the resulting 
non-axisymmetric feature would have a low amplitude. This could explain 
some of the putative spiral features seen in  LSB galaxies. 
If the disc
self-gravity is important then the resonance lasts much longer and this results 
in a regular swing amplification process as discussed in Toomre (1981) 
where the amplitude (the fractional increase in the surface density, or $MAF$) 
will be much higher. Such high amplitude swing amplified features are 
suppressed in the LSBs due to the low disc surface density and the 
dominant dark matter halo as shown in this paper.

We caution that even the LSB galaxy, UGC 7321, that we have studied, 
may possibly show feeble spiral structure. It is seen edge-on with an 
inclination of 88$^0$ (Matthews 2000), so we do not know if it would 
exhibit spiral structure if it were to be seen face-on. It appears to 
have only nominal dust content and patchy star formation (Matthews 2000), 
but both are present at low levels, and so are not totally suppressed.

Alternatively, bars or a central oval feature can trigger a global m=2 
spiral pattern (Binney \& Tremaine 1987), or the spiral arms could arise 
due to manifold-driven trajectories in a barred potential (Athanassoula 2012). 
It is interesting that the few LSB galaxies which show a well-defined though weak 
spiral pattern do have this starting at the end of a bar or an oval, as seen 
in F577-V1 (McGaugh et al. 1995), or in F568-1 (Fuchs 2008). Thus such a 
two-armed spiral pattern could have its origin due to the bar. The question 
then is what gave rise to the bar in the first place despite the halo
dominance in the LSB galaxies. 

A tidal encounter could trigger short-lived spiral 
features in the LSB galaxies, as it does in the HSB galaxies (e.g., Noguchi 1987), 
except that disc response would be lower due to the high $Q$ values and hence the spiral 
features would be weaker.

Some authors have argued that discs in LSBs should have higher density 
to explain the observed spiral features. We discuss below that the 
derivation of these results is problematic. Fuchs (2008) assumes m=2, and 
$X=2$ as seen for high amplification (Toomre 1981) to be applicable, and 
then calculates the disc surface density and finds this to be high. We caution 
that this result is only be valid for those few LSBs which have an m=2 
pattern and is not generally applicable. 
Saburova, Bizyaev \& Zasov 2011, and Saburova \& Zasov (2013) assume 
the disc in the LSBs to be in a marginal stability with $Q = 1$ and then use
this to obtain the stellar surface density. However, 
 the true $Q$ values for the LSBs are $>>1$ (see our 
Table 1). Thus artificially imposing a condition of $Q=1$ as done by Saburova et al. gives a spuriously 
high value of the disc surface density.

In summary, our stability analysis in this paper shows that strong spiral 
features are not likely to be seen in the LSBs due to the halo dominance.
 However, as argued above, the LSBs could still 
 support weak spiral structure due to the kinematic resonance 
between shear and epicyclic motion even without much help from disc self- 
gravity. This could explain the presence of occasional, weak 
spiral structure seen in some LSB galaxies.

\subsection{Generalization to other LSB galaxies}
We have presented results pertaining to a specific LSB galaxy, UGC 7321, 
since the various input parameters are known observationally for this galaxy.
Generalizing the results from this one case to all LSBS may seem a little 
far-fetched. However, we note that the two main physical properties that 
we find to be responsible for suppression of the instabilities in this galaxy 
are the low disc surface density and the dark matter halo that is dominant 
from inner regions. These two features are generally valid for all the LSBs 
(e.g., de Blok \& McGaugh 1996, Bothun et al. 1997, de Blok \& McGaugh 1997). Hence we expect our
conclusion about the suppression of gravitational instabilities to be valid in general
for all the LSB galaxies.

While for several LSBs, rotation curves (e.g., de Blok, McGaugh \&  van der Hulst 1996) and $HI$ surface density (e.g., de Blok et al. 1996, Pickering et al. 1997) have been obtained, the values of velocity dispersion are not known, these are needed so that the idea be applied to other LSBs as well.

\section{Conclusions}
To summarise, we have studied the axisymmetric and non-axisymmetric local, linear  perturbations for a LSB galaxy, UGC 7321, treating the galactic disc as a system of stars-alone and then as a gravitationally coupled system of stars and gas embedded in a rigid  dark matter halo. We show that the galaxy
 is quite stable against both cases when the effect of the dark matter halo is incorporated while studying its dynamics. 
The inclusion of the massive halo results in a higher undisturbed rotational velocity and hence higher $Q$ values. The low disc surface density also contributes to a high $Q$ value. This results in the disc being stable against axisymmetric perturbations, and also it leads to a suppression of swing amplification of non-axisymmetric perturbations. This can explain the lack of star formation and the absence of strong spiral structure in this galaxy.
 Thus the low disc surface density and the dominant
 dark matter halo together have a profound effect on stabilizing the disc against both axisymmetric and non-axisymmetric perturbations in 
LSB galaxies.

\bigskip

\noindent {\bf ACKNOWLEDGEMENTS}

\medskip

We thank the anonymous referee for constructive comments which have greatly improved the paper.

\bigskip

\noindent {\bf References}

\medskip

\noindent  Athanassoula, E., 2012, MNRAS, 426, L46

\noindent Banerjee, A.,  Jog, C.J., 2013, MNRAS, 431, 582B

\noindent Banerjee, A.,  Matthews, L.D.,  Jog, C.J., 2010, New Astronomy, 15, 89

\noindent Bothun, G., Impey, C., McGaugh, S., 1997, PASP, 109, 7453

\noindent Binney, J., Tremaine, S., 1987, Galactic Dynamics. Princeton
Univ. Press, Princeton, NJ

\noindent  Bournaud, F., Combes, F., Jog, C.J., Puerari, I. 2005, A \& A, 438, 507

\noindent Combes, F., 2012, New Astro. Rev., 46, 755

\noindent de Blok, W.J.G., McGaugh, S.S., 1996, ApJ, 469, L89

\noindent de Blok, W.J.G., McGaugh, S.S., 1997, MNRAS, 290, 533

\noindent de Blok, W.J.G., McGaugh, S.S., Rubin, V.C., 2001, AJ, 122, 2396

\noindent de Blok, W.J.G., McGaugh, S.S., van der Hulst, J.M., 1996, MNRAS, 283, 18

\noindent Fuchs, B., 2008, AN, 329, 916

\noindent Goldreich, P., Lynden-Bell, D., 1965, MNRAS,130, 125 (GLB)

\noindent Impey, C., Bothun, G.D., 1997, ARAA, 35, 267

\noindent Jog, C.J., 1992, ApJ, 390, 378

\noindent Jog, C.J., 1996, MNRAS, 278, 209

\noindent Jog, C.J., 2012, in "Recent advances is star formation: Observations and theory", Astronomical Society of India Conference Series, Eds. A. Subramaniam and S. Anathpindika, Vol. 4, p. 145

\noindent Jog, C.J., Solomon, P.M., 1984, ApJ, 276, 114J

\noindent Kennicutt, R.C., 1989, ApJ, 344, 685

\noindent Matthews, L.D., 2000, AJ, 120, 1764

\noindent Matthews, L.D., Gallagher, J.S., 1997, AJ, 114, 1899

\noindent McGaugh, S.S., Schombert, J.M., Bothun, G.D., 1995, AJ, 109, 2019

\noindent Mihos, J.C., McGaugh, S.S., de Blok, W.J.G., 1997, 477, L79

\noindent Mo, H.J., McGaugh, S.S., Bothun, G.D. 1994, MNRAS, 267, 129

\noindent Narayan, C.A., Jog, C.J. 2002, A\&A, 394, 89

\noindent Noguchi, M., 1987, MNRAS, 228, 635

\noindent O'Neil, K., Schinnerer, E., Hofner, P., 2003, ApJ, 588, 230

\noindent Pickering, T.E., Impey, C.D., van Gorkom, J.H.,  Bothun, G.D., 1997, AJ, 114, 1858

\noindent Rix, H.-W., Zaritsky, D., 1995, ApJ, 447, 82

\noindent  Rosenbaum, S.D., Krusch, E., Bomans, D.J., Dettmar, R.-J. 2009, A\&A, 504, 807

\noindent Saburova, A.S., Bizyaev, D.V., Zasov, A.V., 2011, Astronomy Letters, 37, 751

\noindent Saburova, A.S., Zasov, A.S., 2013, AN,
334, 785

\noindent Schombert, J. M., Bothun, G. D., Impey, C. D., Mundy, L. G., 1990, AJ, 100, 1523

\noindent Schombert, J. M., Maciel, T., McGaugh, S.S., 2011, Adv. Astr., 
Article ID 143698

\noindent  Sellwood, J.A., Carlberg, R.G., 1984, ApJ, 282, 61

\noindent Toomre, A., 1964, ApJ, 139, 1217

\noindent Toomre, A., 1981, in Structure and Dynamics of Normal Galaxies, 
      ed. S.M. Fall \& D. Lynden-Bell (Cambridge: Cambridge Univ. Press), pg. 111

\noindent Uson, J.M.,  Matthews, L.D., 2003, AJ, 125, 2455

\noindent  van der Hulst, J. M., Skillman, E. D., Smith, T. R., Bothun, G. D., McGaugh, S. S., de Blok, W. J. G.,
          1993, AJ, 106, 548

\end{document}